\documentstyle[12pt, axodraw]{article}

\input amssym.tex

\begin{document}

\renewcommand{\a}{\alpha}
\renewcommand{\b}{\beta}
\renewcommand{\d}{\delta}         \newcommand{\D}{\Delta}
\newcommand{\ve}{\varepsilon}
\newcommand{\eps}{\epsilon}
\newcommand{\k}{\kappa}
\newcommand{\ld}{\lambda}        \newcommand{\LD}{\Lambda}
\newcommand{\om}{\omega}         \newcommand{\OM}{\Omega}
\newcommand{\p}{\psi}             \newcommand{\PS}{\Psi}
\newcommand{\ro}{\rho}
\newcommand{\s}{\sigma}           \renewcommand{\S}{\Sigma}
\newcommand{\th}{\theta}         \newcommand{\T}{\Theta}
\newcommand{\f}{{\phi}}           \newcommand{\F}{{\Phi}}
\newcommand{\vf}{{\varphi}}
\newcommand{\y}{{\upsilon}}       \newcommand{\Y}{{\Upsilon}}
\newcommand{\z}{\zeta}
\newcommand{\X}{\Xi}
\newcommand{\cA}{{\cal A}}
\newcommand{\cB}{{\cal B}}
\newcommand{\cC}{{\cal C}}
\newcommand{\cD}{{\cal D}}
\newcommand{\cE}{{\cal E}}
\newcommand{\cF}{{\cal F}}
\newcommand{\cG}{{\cal G}}
\newcommand{\cH}{{\cal H}}
\newcommand{\cI}{{\cal I}}
\newcommand{\cJ}{{\cal J}}
\newcommand{\cK}{{\cal K}}
\newcommand{\cL}{{\cal L}}
\newcommand{\cM}{{\cal M}}
\newcommand{\cN}{{\cal N}}
\newcommand{\cO}{{\cal O}}
\newcommand{\cP}{{\cal P}}
\newcommand{\cQ}{{\cal Q}}
\newcommand{\cS}{{\cal S}}
\newcommand{\cR}{{\cal R}}
\newcommand{\cT}{{\cal T}}
\newcommand{\cU}{{\cal U}}
\newcommand{\cV}{{\cal V}}
\newcommand{\cW}{{\cal W}}
\newcommand{\cX}{{\cal X}}
\newcommand{\cY}{{\cal Y}}
\newcommand{\cZ}{{\cal Z}}
\newcommand{\hA}{{\widehat A}}
\newcommand{\hB}{{\widehat B}}
\newcommand{\hC}{{\widehat C}}
\newcommand{\hD}{{\widehat D}}
\newcommand{\hE}{{\widehat E}}
\newcommand{\hF}{{\widehat F}}
\newcommand{\hG}{{\widehat G}}
\newcommand{\hH}{{\widehat H}}
\newcommand{\hI}{{\widehat I}}
\newcommand{\hJ}{{\widehat J}}
\newcommand{\hK}{{\widehat K}}
\newcommand{\hL}{{\widehat L}}
\newcommand{\hM}{{\widehat M}}
\newcommand{\hN}{{\widehat N}}
\newcommand{\hO}{{\widehat O}}
\newcommand{\hP}{{\widehat P}}
\newcommand{\hQ}{{\widehat Q}}
\newcommand{\hS}{{\widehat S}}
\newcommand{\hR}{{\widehat R}}
\newcommand{\hT}{{\widehat T}}
\newcommand{\hU}{{\widehat U}}
\newcommand{\hV}{{\widehat V}}
\newcommand{\hW}{{\widehat W}}
\newcommand{\hX}{{\widehat X}}
\newcommand{\hY}{{\widehat Y}}
\newcommand{\hZ}{{\widehat Z}}
\newcommand{\Ha}{{\widehat a}}
\newcommand{\Hb}{{\widehat b}}
\newcommand{\Hc}{{\widehat c}}
\newcommand{\Hd}{{\widehat d}}
\newcommand{\He}{{\widehat e}}
\newcommand{\Hf}{{\widehat f}}
\newcommand{\Hg}{{\widehat g}}
\newcommand{\Hh}{{\widehat h}}
\newcommand{\Hi}{{\widehat i}}
\newcommand{\Hj}{{\widehat j}}
\newcommand{\Hk}{{\widehat k}}
\newcommand{\Hl}{{\widehat l}}
\newcommand{\Hm}{{\widehat m}}
\newcommand{\Hn}{{\widehat n}}
\newcommand{\Ho}{{\widehat o}}
\newcommand{\Hp}{{\widehat p}}
\newcommand{\Hq}{{\widehat q}}
\newcommand{\Hs}{{\widehat s}}
\newcommand{\Hr}{{\widehat r}}
\newcommand{\Ht}{{\widehat t}}
\newcommand{\Hu}{{\widehat u}}
\newcommand{\Hv}{{\widehat v}}
\newcommand{\Hw}{{\widehat w}}
\newcommand{\Hx}{{\widehat x}}
\newcommand{\Hy}{{\widehat y}}
\newcommand{\Hz}{{\widehat z}}
\newcommand{\deff}{\,\stackrel{\rm def}{\equiv}\,}
\newcommand{\lra}{\longrightarrow}
\newcommand{\ra}{\,\rightarrow\,}
\def\limar#1#2{\,\raise0.3ex\hbox{$\longrightarrow$\kern-1.5em\raise-1.1ex
\hbox{$\scriptstyle{#1\rightarrow #2}$}}\,}
\def\limarr#1#2{\,\raise0.3ex\hbox{$\longrightarrow$\kern-1.5em\raise-1.3ex
\hbox{$\scriptstyle{#1\rightarrow #2}$}}\,}
\def\limlar#1#2{\ \raise0.3ex
\hbox{$-\hspace{-0.5em}-\hspace{-0.5em}-\hspace{-0.5em}
\longrightarrow$\kern-2.7em\raise-1.1ex
\hbox{$\scriptstyle{#1\rightarrow #2}$}}\ \ }
\newcommand{\limm}[2]{\lim_{\stackrel{\scriptstyle #1}{\scriptstyle #2}}}
\newcommand{\wt}{\widetilde}
\newcommand{\os}{{\otimes}}
\newcommand{\da}{{\dagger}}
\newcommand{\stimes}{\times\hspace{-1.1 em}\supset}
\def\h{\hbar}
\newcommand{\ih}{\frac{\i}{\h}}
\newcommand{\exx}[1]{\exp\left\{ {#1}\right\}}
\newcommand{\ord}[1]{\mbox{\boldmath{$\cO$}}\left({#1}\right)}
\newcommand{\one}{{\leavevmode{\rm 1\mkern -5.4mu I}}}
\newcommand{\Z}{Z\!\!\!Z}
%
\newcommand{\Ibb}[1]{ {\rm I\ifmmode\mkern
            -3.6mu\else\kern -.2em\fi#1}}
\newcommand{\ibb}[1]{\leavevmode\hbox{\kern.3em\vrule
     height 1.2ex depth -.3ex width .2pt\kern-.3em\rm#1}}
\newcommand{\N}{{\Ibb N}}
\newcommand{\R}{{\Ibb R}}
\newcommand{\HH}{{\Ibb H}}
\newcommand{\rational}{{\kern .1em {\raise .47ex
\hbox{$\scripscriptstyle |$}}
    \kern -.35em {\rm Q}}}
\newcommand{\bm}[1]{\mbox{\boldmath${#1}$}}
\newcommand{\intf}{\int_{-\infty}^{\infty}\,}
\newcommand{\LL}{\cL^2(\R^2)}
\newcommand{\LLS}{\cL^2(S)}
\newcommand{\Ree}{{\cal R}\!e \,}
\newcommand{\Imm}{{\cal I}\!m \,}
\newcommand{\tr}{{\rm {Tr} \,}}
\newcommand{\er}{{\rm{e}}}
\renewcommand{\i}{{\rm{i}}}
\newcommand{\divv}{{\rm {div} \,}}
\newcommand{\id}{{\rm{id}\,}}
\newcommand{\ad}{{\rm{ad}\,}}
\newcommand{\Ad}{{\rm{Ad}\,}}
\newcommand{\const}{{\rm{\,const\,}}}
\newcommand{\rank}{{\rm{\,rank\,}}}
\newcommand{\diag}{{\rm{\,diag\,}}}
\newcommand{\sign}{{\rm{\,sign\,}}}
\newcommand{\pa}{\partial}
\newcommand{\pad}[2]{{\frac{\partial #1}{\partial #2}}}
\newcommand{\padd}[2]{{\frac{\partial^2 #1}{\partial {#2}^2}}}
\newcommand{\paddd}[3]{{\frac{\partial^2 #1}{\partial {#2}\partial {#3}}}}
\newcommand{\der}[2]{{\frac{{\rm d} #1}{{\rm d} #2}}}
\newcommand{\derr}[2]{{\frac{{\rm d}^2 #1}{{\rm d} {#2}^2}}}
\newcommand{\fud}[2]{{\frac{\delta #1}{\delta #2}}}
\newcommand{\fudd}[2]{{\frac{\d^2 #1}{\d {#2}^2}}}
\newcommand{\fuddd}[3]{{\frac{\d^2 #1}{\d {#2}\d {#3}}}}
\newcommand{\dpad}[2]{{\displaystyle{\frac{\partial #1}{\partial #2}}}}
\newcommand{\dfud}[2]{{\displaystyle{\frac{\delta #1}{\delta #2}}}}
\newcommand{\dd}{\partial^{(\ve)}}
\newcommand{\ddd}{\bar{\partial}^{(\ve)}}
\newcommand{\dfrac}[2]{{\displaystyle{\frac{#1}{#2}}}}
\newcommand{\dsum}[2]{\displaystyle{\sum_{#1}^{#2}}}
\newcommand{\dint}{\displaystyle{\int}}
\newcommand{\dg}{\!\not\!\partial}
\newcommand{\vg}[1]{\!\not\!#1}
\def\<{\langle}
\def\>{\rangle}
\def\lgl{\langle\langle}
\def\rgr{\rangle\rangle}
\newcommand{\bra}[1]{\left\langle {#1}\right|}
\newcommand{\ket}[1]{\left| {#1}\right\rangle}
\newcommand{\vev}[1]{\left\langle {#1}\right\rangle}
\newcommand{\be}{\begin{equation}}
\newcommand{\ee}{\end{equation}}
\newcommand{\bn}{\begin{eqnarray}}
\newcommand{\en}{\end{eqnarray}}
\newcommand{\bnn}{\begin{eqnarray*}}
\newcommand{\enn}{\end{eqnarray*}}
\newcommand{\e}{\label}
\newcommand{\nbr}{\nonumber\\[2mm]}
\newcommand{\r}[1]{(\ref{#1})}
\newcommand{\refp}[1]{\ref{#1}, page~\pageref{#1}}
\renewcommand {\theequation}{\thesection.\arabic{equation}}
\renewcommand {\thefootnote}{\fnsymbol{footnote}}
\newcommand{\qq}{\qquad}
\newcommand{\qqq}{\quad\quad}
\newcommand{\biz}{\begin{itemize}}
\newcommand{\eiz}{\end{itemize}}
\newcommand{\ben}{\begin{enumerate}}
\newcommand{\een}{\end{enumerate}}
\def\nc{noncommutative}
\def\ncy{noncommutativity}
\def\com{commutative}
\def \simlt{\stackrel{<}{{}_\sim}}
\thispagestyle{empty}
\begin{flushright}
HIP-2003-32/TH\\
\end{flushright}
\vspace{2cm}
\begin{center}

{\Large{\bf{Tree Unitarity and Partial Wave Expansion\\
in Noncommutative Quantum Field Theory}}}
\vskip .7cm
{\bf{\large{M. Chaichian$^\dagger$, C. Montonen$^{\dagger\dagger}$
\ \ and \ \ A. Tureanu$^\dagger$}}}

{\it $^\dagger$High Energy Physics Division, Department of
Physical Sciences,
University of Helsinki\\
\ \ {and}\\
\ \ Helsinki Institute of Physics,
P.O. Box 64, FIN-00014 Helsinki, Finland\\
$^{\dagger\dagger}$Division of Theoretical Physics, Department of
Physical Sciences, University of Helsinki,\\
P.O. Box 64, FIN-00014 Helsinki, Finland}

\setcounter{footnote}{0}

\vspace{2cm}
{\bf Abstract}

\end{center}  

The validity of the tree-unitarity criterion for scattering amplitudes on the \nc\ space-time is considered,
as a condition that can be used to shed  light on the problem of unitarity violation in \nc\ quantum field
theories when time is \nc. The unitarity constraints on the partial wave amplitudes in the \nc\ space-time are
also derived. 

\vskip .3cm

\newpage
\section{Introduction}
\setcounter{equation}{0}

Recently, quantum field theories on noncommutative (NC) space-time have received a lot of attention, after it
was discovered that, in some cases, they emerge naturally as low-energy limits from string theory with an antisymmetric
background field \cite{SW}. On a noncommutative analog of the Minkowski space, the coordinates satisfy
non-trivial commutation relations:
\be
[\hat{x_\mu},\hat{x}_\nu]=i\theta_{\mu\nu}\ ,
\ee
where $\theta_{\mu\nu}$ is a constant antisymmetric matrix of dimension (length)$^2$. The inherent
non-locality and the violation of Lorentz invariance in NC QFT are the main causes which lead to some peculiar
features in the case of \nc\ models. 

The question of unitarity of theories with  time-space \ncy\ ($\theta_{0i}\neq 0$) is a topical one in NC
QFT. It was first shown in \cite{Gomis} that such theories are not perturbatively unitary when naive Feynman
rules are used, but also that they
cannot be obtained as low-energy limits from the underlying string theory (see also \cite{discrete} for a
study of the violation of unitarity on compact space-time). The subject was approached later
again in \cite{Bahns}, in the light of the Yang-Feldman equation \cite{Yang}, thereby arriving at a manifestly
Hermitian solution (hence unitary theory with $\theta_{0i}\neq 0$). The study was further pursued in 
\cite{{Liao1},{Liao3}}, where the Wick contraction theorem was adapted to the case when time does not
commute with space, hence the time-ordering procedure does not commute with the star multiplication. As a result,
a \nc\ extension of the time-ordered perturbation theory (TOPT) was formulated, 
which gives the same results as the
standard procedure (in terms of ordinary Feynman propagators, introduced in \cite{Filk}) for $\theta_{0i}= 0$, but differs 
from it in the case when $\theta_{0i}\neq 0$. It is claimed, and cheked in the few lowest orders, that this
formulation leads to theories which are
perturbatively unitary \cite{Liao1}. However, NC QED treated according to the TOPT prescription shows
 a "surprising result" \cite{Liao3} regarding
the high-energy behaviour of the two-body cross-sections: it yields cross-sections, calculated
in the lowest order perturbation theory, exhibiting a growth linear in $s$\footnote{In \cite{Liao3} it is
stated that the same phenomenon occurs when the mass of the exchanged particle is much less than the NC energy
scale and the center-of-mass energy. However, a straightforward calculation (see eq. (\ref{scat_ampl})) shows that this is not true for
the NC $\phi^3$ scalar theory, in which case the two-body cross-section tends to 0 when
$E_{CM}\rightarrow\infty$, although not as fast as when it is computed in the standard ("covariant")
perturbation theory.}. 
It is therefore of interest to apply 
other criteria, such as the tree-unitarity conditions and see whether they are
violated. The fact that time-space NC quantum field theories, in addition to the
impossibility of their being obtained from the string theory \cite{{SW},{Gomis}}, violate causality on both the macro-
and microscopic levels \cite{{Seiberg},{LAG},{CNT}}, gives reasons to expect that this could be the case.

The scope of this letter is two-fold: on the one hand, we would like to check if the theories with time-space
\ncy, treated according to the TOPT prescriptions, satisfy the tree-unitarity criterion \cite{{tikt1},{ls}}.
Such a consideration would be interesting, since in the past the requirement of mere tree-unitarity was 
successful in distinguishing among different models with respect to their unitarity/renormalizability
\cite{{tikt1},{ls}}. One could hope that the same merit would hold also in the case of NC theories. 

On the other hand, 
we would like to derive a partial wave expansion and unitarity constraints
on the partial wave amplitudes in \nc\ space (actually, in any nonisotropic space-time), as tools
(together with the analyticity of the scattering amplitude and the dispersion relations) for the derivation of
bounds on the cross-section and the amplitudes themselves, analogous to the celebrated Froissart-Martin bound
\cite{{Froissart}, {Martin}} in the usual QFT.

Notation: In the following we shall denote $\epsilon_i=\theta_{0i}$ and 
$\beta_i=\frac{1}{2}\epsilon_{ijk}\theta_{jk}$.

\section{Tree unitarity}

To begin with, we shall recall the concept of tree unitarity \cite{tikt1}.
The unitarity of the $S$-matrix, written in the familiar way with respect to the transition amplitude
\cite{ELOP}
\be\label{s-matrix}
S=1+iT,
\ee
implies the following condition on the transition amplitude:
\be\label{t-matrix}
T-T^\dagger=iTT^\dagger=iT^\dagger T. 
\ee
The on-shell transition amplitude between the initial state $|i\rangle$ and the final state $|f\rangle$ is
\be
\langle f|T|i\rangle=(2\pi)^4\delta(P'-P)\langle f|A|i\rangle\ ,
\ee
where $P, P'$ are the initial and final four-momenta. We assume that the energy-momentum
dispersion relation still takes the form $E=\sqrt{\vec k^2+m^2}$ in the \nc\ case. From (\ref{t-matrix}) it follows that
the $A$-matrix elements satisfy the unitarity relation:
\be
-\frac{i}{2}(\langle f|A|i\rangle-\langle i|A|f\rangle^*)=\frac{1}{2}\sum_n(2\pi)^{4-3n}\int\frac{d^3k_1}{2k^0_{1}}\cdots\frac{d^3k_n}{2k^0_{n}}
\delta(\sum\ k_i-P)\langle k_1\cdots k_n|A|f\rangle^*\langle k_1\cdots k_n|A|i\rangle\ .
\ee
Denote by $A_{n\rightarrow N-n}$ an $A$-matrix for $n$ incoming particles and $N-n$ outgoing particles. In the
center-of-mass frame, one chooses {\it fixed} values for the incoming and outgoing momenta, so that for given
values of these "fixed variables" each four-momentum $p_i$ grows as $E$, as the total center-of-mass energy
($E$) approaches infinity. A field theory will be called {\it tree unitary} if in the tree approximation all
amplitudes $A_{n\rightarrow N-n}$ grow at most like $E^{4-N}$ as $E\rightarrow \infty$. In other words, if at high
energies $A_{n\rightarrow N-n}\sim E^\beta$, then the requirement of tree unitarity can be expressed in the form
\be\label{tree}
\beta\leq 4-N\ .
\ee

In the \nc\ quantum field theory the crossing symmetry is still holds, 
but it is lost when one goes to a specific reference frame and specific initial and final states (as required
by the tree-unitarity criterion),so that we need to check separately if the tree
unitarity is fulfilled for the $s$- and $t$-channels.

We shall begin with the $s$-channel. One typical tree-level scattering amplitude was obtained 
in the first paper of \cite{Liao1}, for a two-by-two scattering
$\pi(p_1)\pi(p_2)\rightarrow\chi(p_3)\chi(p_4)$ through the cubic scalar interactions defined by the
Lagrangean $L_{int}=-g_\pi\pi\star\sigma\star\pi-g_\chi\chi\star\sigma\star\chi$
 (the fields were taken to be
non-identical in order to reduce the number of channels to one). The expression for the $2\rightarrow 2
$ scattering amplitude, in the $s$-channel and in the
center-of-mass frame, can be cast into the form:
\bn\label{scat_ampl}
A_{2\rightarrow 2}^s(\vec{p};\vec{p'})&=&\frac{2g_\pi g_\chi}{s-m^2_\sigma}\sum_{\ld=\pm 1}
\{\cos[m_\sigma(\tilde p_o+\ld\tilde p'_0)]\cos[\frac{\sqrt s}{2}(\tilde p_o+\ld\tilde p'_0)]\cr
&+&\frac{\sqrt s}{m_\sigma}\sin[m_\sigma(\tilde p_o+\ld\tilde p'_0)]
\sin[\frac{\sqrt s}{2}(\tilde p_o+\ld\tilde p'_0)]\}\ ,
\en
where $\tilde p_0=\theta_{0i}p^i=\vec{\epsilon}\cdot\vec{p}$ and $m_\sigma$ is the mass of the 
$s$-channel scalar particle\footnote {To prove the affirmation of the previous footnote, one can plug the
expression (\ref{scat_ampl}) into the formula of the differential cross-section calculated in CMS for external
particles with equal mass, i.e. $(\frac{d\sigma}{d\Omega})_{CM}=\frac{|A|^2}{64\pi^2E^2_{CM}}$. It is clear
that at high energies, the differential cross-section behaves at most like $\frac{1}{s^2}$.}. 
The second term in the brackets, proportional to $\sqrt s=E$, is an element of novelty in the TOPT as compared with
the usual "covariant" approach. 
However, when we take the limit $E\rightarrow\infty$, the $2\rightarrow 2$ amplitude ($N=4$) behaves like 
$\frac{E}{E^2}=E^{-1}$, thus fulfilling the tree-unitarity criterion, which requires it to grow not faster than 
$E^{(N-4)}=E^0$.

In order to be able to appreciate if the tree-unitarity criterion is satisfied in general, we shall move
further to the 5-point amplitude $A_{2\rightarrow 3}^s$.

\begin{figure}
\begin{center}
\begin{picture}(250,150)(0,0)
\SetOffset(35,0)
\ArrowLine(0,0)(30,40)\ArrowLine(0,80)(30,40)
\ArrowLine(30,40)(90,40)\ArrowLine(90,40)(150,40)
\ArrowLine(90,40)(120,80)
\ArrowLine(150,40)(180,80)\ArrowLine(150,40)(180,0)
\Text(0,20)[]{$k_1$}\Text(155,20)[]{$k_4$}
\Text(0,60)[]{$k_2$}\Text(155,60)[]{$k_3$}
\Text(60,30)[]{$p$}\Text(120,30)[]{$q$}\Text(80,60)[]{$k_5$}
\end{picture}
\end{center}
\caption{Diagram corresponding to $A_{2\rightarrow 3}^s$.}
\end{figure}
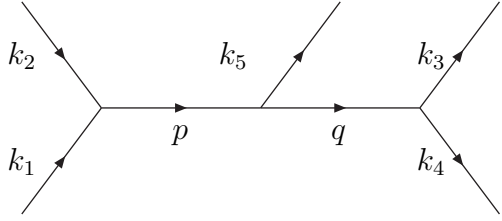

The expression of the amplitude, according to the TOPT prescription, is:
\bn
A_{2\rightarrow 3}^s&\sim& g_\pi g^2_\chi\delta(E_1+E_2-E_3-E_4-E_5)\sum_{\lambda_1,\lambda_2=\pm1}
\int\frac{d^3p}{(2\pi)^3 2\omega_p}
\frac{d^3q}{(2\pi)^3 2\omega_q}\cr
&\times& (2\pi^3)\delta(\vec k_1+\vec k_2+\vec p)(2\pi^3)\delta(\vec p+\vec q+\vec k_5)
(2\pi^3)\delta(\vec q+\vec k_3+\vec k_4)\\
&\times&\frac{[e^{-i(k_{1+},-p_{\lambda_1},k_{2+})}+(k_1\rightarrow k_2)]
[e^{-i(k_{5-},p_{\lambda_1},-q_{\lambda_2})}+(q\rightarrow k_5)]
[e^{-i(q_{\lambda_2},k_{3-},k_{4-})}+(q\rightarrow k_4)]}{[\lambda_1(E_1+E_2)-\omega_p+i\epsilon]
[-\lambda_2(E_3+E_4)-\omega_q+i\epsilon]}\ .\nonumber
\en

A typical term of the amplitude is of the form:
\be
\frac{1}{\omega_p\omega_q}\sum_{\lambda_1,\lambda_2=\pm1}\frac{e^{i\lambda_1 a +\lambda_2 b+ c}}
{(\lambda_1 E-\omega_p)(-\lambda_2
E'-\omega_q)}\ ,
\ee
where $a$, $b$ and $c$ are factors depending on the momenta of the particles involved in the interaction 
and on 
the \ncy\ parameter $\theta_{\mu\nu}$ (in such a way that $a=b=c=0$ for $\theta=0$), $E=E_1+E_2$ and 
$E'=E_3+E_4$. Performing the summation over $\lambda$'s, 
one obtains:
\bn
\frac{8}{(E^2-\omega_p^2)(E'^2-\omega_q^2)}[(\cos a\cos b-\frac{EE'}{\omega_p\omega_q}\sin a\sin b)\cos c\cr
-(\frac{E}{\omega_p}\sin a\cos b-\frac{E'}{\omega_q}\sin b\cos a)\sin c]\ .
\en

In the center-of-mass frame $E^2=(k_1+k_2)^2=(k_3+k_4+k_5)^2$ and we shall fix the outgoing momenta (in the
spirit of the tree-unitarity criterion) so that $|\vec k_3|=|\vec k_4|=|\vec k_5|$. Assuming for simplicity
the equality of all the masses of the particles involved in the interaction, the following expression is
obtained, for this {\it specific} phase-space configuration, in terms of the center-of-mass energy $E$:
\be
\frac{24}{E^2(E^2-m^2)}[(\cos a\cos b-\frac{2E}{m}\sin a\sin b)\cos c
-(\frac{E}{m}\sin a\cos b-2\sin b\cos a)\sin c]\ .
\ee
The $E$-dependence of $a$, $b$ and $c$, which is of polynomial form, is not relevant for the high-energy
behaviour, as the sine and cosine functions do not have a limit when their arguments are polynomials in $E$ 
for $E\rightarrow\infty$, but still they are
bounded in the interval $[-1,1]$. It becomes clear that for high energies, the typical term of the 5-point
amplitude $A_{2\rightarrow 3}^s$ behaves like
$$
\frac{E}{E^4}=E^{-3}\ .
$$
According to \cite{tikt1}, the 5-point amplitude should not grow faster then $E^{4-N}=E^{-1}$. Obviously, this requirement if fulfilled by the amplitude $A_{2\rightarrow 3}^s$.

We expect
that, in the $s$-channel, the tree-amplitudes $A_{2\rightarrow N-2}^s\sim s^{\beta/2}$, with $N>5$, will behave well at high energies,
so that $\beta<(4-N)$.

We shall now consider the tree-unitarity criterion in the $t$-channel, in which case, for a {\it fixed
configuration}, at high energies, $t\sim s=E^2$. We have computed, according to TOPT prescriptions, 
the $2\rightarrow 2$ scattering amplitude in the
$t$-channel, for an interaction Lagrangean of the form
$L_{int}=-g(\pi\star\sigma\star\chi+\chi\star\sigma\star\pi)$, i.e.:
\be
A_{2\rightarrow 2}^t\sim g^2[\frac{2\cos(k_{1+},-q_+,-k_{3+})2\cos(k_{2+},q_+,
-k_{4+})}{2\omega_q(E_1-E_3-\omega_q+i\epsilon)}+\frac{2\cos(k_{1+},-q_-,-k_{3+})2\cos(k_{2+},q_-,
-k_{4+})}{2\omega_q(E_2-E_4-\omega_q+i\epsilon)}]\ .
\ee
In the center-of-mass frame, and taking for simplicity $m_\pi=m_\chi=m$, 
we obtained:
\bn
A_{2\rightarrow 2}^t&\sim&\frac{2g^2}{t-m_\sigma^2}[\cos(\sqrt{m_\sigma-t}\theta_{0i}(k_1+k_3)^i)
\cos(\theta_{ij}k_1^ik_3^j)+\cos(\frac{1}{2}\sqrt s\theta_{0i}(k_1-k_3)^i)]\ .
\en
In this case, the high-energy behaviour is governed by the first factor (as the cosines are bounded when
$s\rightarrow\infty$) and is the same like in the commutative case. The amplitudes with more legs will show
the same similarity with the commutative case at high energies, and we can conclude that they will satisfy the
tree-unitarity criterion.

\section{Partial wave expansion}

In the commutative case, due to the rotational invariance, the $2\rightarrow 2$ scattering amplitude depends on two variables: $s$ and $t$, 
i.e. the squared center-of-mass energy and the
squared transferred momentum or, equivalently, $s$ and $\cos\theta$, with $\theta$ being the center-of-mass scattering angle. 

The partial wave amplitudes are defined by the expansion in Legendre polynomials
\cite{ELOP}:
\bn\label{pw_cc}
A(s,\cos\theta)&=&\sum_{l=0}^\infty (2l+1)a_l(s)P_l(\cos\theta),\cr
a_l(s)&=&\frac{1}{2}\int_{-1}^1d(\cos\theta)P_l(\cos\theta)A(s,\cos\theta),
\en
where $A(s,\cos\theta)\equiv {\cal A}(s,t)$ is the scattering amplitude in terms of the Mandelstam variables $s$ and
$t=-\frac{s}{2}(1-\cos\theta)$ (for the equal-mass case).

In the \nc\ case the rotational invariance is lost and as a result the number of independent angular variables
is increased. For the general case of space-time \ncy, $\theta_{\mu\nu}$ defines a plane through the vectors
$\epsilon_i=\theta_{0i}$ and $\beta_i=\frac{1}{2}\epsilon_{ijk}\theta_{jk}$. The only symmetry left is then a
reflection in this plane. The situation is thus close to a fully anisotropic (but translationally invariant)
background, and we treat this general case in the following. The results are then generally applicable to
scattering in completely anisotropic media. With respect to arbitrarily chosen axes, the
directions of the three-vectors $\vec{p_1}$ and $\vec{p_3}$ are each given by two angles,  $(\theta_{12},
\phi_{12})$ and $(\theta_{34}, \phi_{34})$, respectively.

However, in the case of space-space \ncy, when $\theta_{0i}=0$, i.e. $\vec{\epsilon}=0$ and in the case of
lightlike \ncy, when $\theta^{\mu\nu}\theta_{\mu\nu}=0$ and $\vec{\epsilon}\perp\vec{\beta}$, there are only
three independent angular variables, which can be assumed to be the angles
$\widehat{(\vec{\beta},\vec{p_1})}$, $\widehat{(\vec{\beta},\vec{p_3})}$ and
$\widehat{(\vec{p_1},\vec{p_3})}$. It should be emphasized, however, that only in these latter two cases
(space-space \ncy and lightlike \ncy), a NC field theory can be obtained from the string theory
as the low-energy limit \cite{{SW},{Gomis},{Aharony}}.

\subsection{Unitarity constraint on partial wave amplitudes}

For a 2-particles initial and final states, the on-shell amplitude is:
\be
\langle p_3,p_4|T|p_1,p_2\rangle=(2\pi)^4\delta(p_1+p_2-p_3-p_4)A(\vec{p_1},\vec{p_2};\vec{p_3},\vec{p_4}).
\ee
Next we expand $A(\vec{p_1},\vec{p_2};\vec{p_3},\vec{p_4})$ in partial waves, demanding that the amplitude is
single-valued. The angular dependence will be taken into account
through the spherical harmonics $Y_{lm}(\theta_{12},\phi_{12})$ and $Y_{l'm'}(\theta_{34},\phi_{34})$, while 
the dependence on $s$ will be
accounted for through the partial-wave amplitudes $a_{lm,l'm'}(s)$, i.e.
\be\label{pw}
A(\vec{p};\vec{p'})=4\pi\sum_{l,l',m,m'}a_{lm,l'm'}(s)Y_{lm}(\theta_{12},\phi_{12})Y_{l'm'}(\theta_{34},\phi_{34}).
\ee
(When there is only one preferred direction in space, e.g. $\vec{\epsilon}=0$, $\vec{\beta}\neq0$, invariance
under rotations arround that direction implies that the scattering amplitude does not depend on e.g.
$\phi_{12}$. The expansion in that situation becomes a special case of the general formula (\ref{pw}), with
only terms with $m=0$ surviving.)

The bound can be obtained using the relation 
between the elastic cross-section and scattering amplitude,
\be
\sigma_ {el}=\frac{1}{64\pi^2s}\int d \Omega_{34}|A|^2\ ,
\ee
and the optical theorem for forward scattering (i.e. $p_1=p_3$ and $p_2=p_4$), written in the form
\be\label{opt_th}
Im\ A(s)_{forward}=2\sqrt s\ p\ \sigma_{tot},
\ee
when the two particles in the initial state have equal masses and $\sigma_{tot}$ is the total
cross-section. Then, using the expansion (\ref{pw}), 
the elastic
cross-section becomes:
\be
\sigma_{el}=\frac{1}{4s}\sum_{l_1,l_2,l',m_1,m_2,m'}a_{l_1,m_1,l'm'}(s)a^*_{l_2,m_2,l'm'}(s)
Y_{l_1m_1}(\theta_{12},\phi_{12})Y^*_{l_2m_2}(\theta_{12},\phi_{12}).
\ee
and the r.h.s of (\ref{opt_th}) will be:
\be
Im\ A(s)_{forward}=-2\pi i\sum_{l,l',m,m'}[a_{lm,l'm'}(s)-(-1)^{m+m'}a_{l,-m,l',-m'}(s)]
Y_{lm}(\theta_{12},\phi_{12})
Y^*_{l'm'}(\theta_{12},\phi_{12})\ .
\ee
Taking into account that $\sigma_{el}\leq\sigma_{tot}$, it follows that
\bn\label{exact}
(-i)\sum_{l,l',m,m'}[a_{lm,l'm'}(s)-(-1)^{m+m'}a_{l,-m,l',-m'}(s)]Y_{lm}(\theta_{12},\phi_{12})
Y_{l'm'}(\theta_{12},\phi_{12})\cr
\geq\frac{p}{4\pi\sqrt s}\sum_{l,l',l_1,m,m',m_1}(-1)^{m'}
a^*_{l',-m',l_1,m_1}(s)a_{l,m,l_1,m_1}(s)Y_{lm}(\theta_{12},\phi_{12})
Y_{l'm'}(\theta_{12},\phi_{12})\ .
\en

The expression (\ref{exact}) is an exact unitarity condition on the partial-wave amplitudes. As the sign
between the two sides is an inequality, one can not use the orthonormality property of the spherical
harmonics, because they do not have a definite sign on the whole domain of their arguments.

However, for energies were elastic unitarity is exact, on can obtain approximate unitarity conditions on the
partial-wave amplitudes, but with an equality sign, which will make the situation easier to deal with.

With the following convention for one-particle states:
\bn\label{conv}
\langle p|p'\rangle&=&(2\pi)^3 2p_0\delta(\vec{p}-\vec{p'})\ ,\cr
1&=&\int\frac{d^3p}{2p_0(2\pi)^3}|p\rangle\langle p|,
\en
we can write the {\it elastic} unitarity condition in terms of $A(\vec{p_1},\vec{p_2};\vec{p_3},\vec{p_4})$:
\bn\label{a-unit}
A(\vec{p_1},\vec{p_2};\vec{p_3},\vec{p_4})-A^*(\vec{p_3},\vec{p_4};\vec{p_1},\vec{p_2})&=&
\frac{i}{(2\pi)^2}\int\frac{d^3k_1}{2k_1^0}\frac{d^3k_2}{2k_2^0}\delta(p_1+p_2-k_1-k_2)\cr
&\times&A^*(\vec{p_3},\vec{p_4};\vec{k_1},\vec{k_2})
A(\vec{p_1},\vec{p_2};\vec{k_1},\vec{k_2})\ .
\en
In the center-of-mass frame, where $\vec{p_1}=-\vec{p_2}=\vec{p}$,  $\vec{p_3}=-\vec{p_4}=\vec{p'}$
and $\vec{k_1}=-\vec{k_2}=\vec{k}$, (\ref{a-unit}) becomes:
\bn
A(\vec{p};\vec{p'})-A^*(\vec{p'};\vec{p})
&=&\frac{i}{(2\pi)^2}\frac{1}{8}\int d\Omega_{\vec{k}}\int_0^\infty\frac{k^2dk}{k^2+m^2}
\delta(\sqrt{k^2+m^2}-\sqrt{p^2+m^2})
A^*(\vec{p'};\vec{k})A(\vec{p};\vec{k})\cr
&=&\frac{i}{(2\pi)^2}\frac{1}{8}\int d\Omega_{\vec{k}}\frac{p}{\sqrt{p^2+m^2}}
A^*(\vec{p'};\vec{k})A(\vec{p};\vec{k})\ .
\en
Taking into account that $\sqrt{p^2+m^2}=\frac{\sqrt s}{2}$, one obtains:
\be\label{unitarity}
(-i)[A(\vec{p};\vec{p'})-A^*(\vec{p'};\vec{p})]=
\frac{1}{16\pi^2}\frac{p}{\sqrt s}\int
d\Omega_{\vec{k}}A^*(\vec{p'};\vec{k})A(\vec{p};\vec{k})\ ,
\ee
where $p$ is the magnitude of the three-momentum of the initial particles in the center-of-mass frame.

With the expansion (\ref{pw}), the integral in the r.h.s of (\ref{unitarity}) becomes:
\bn\label{rhs}
&\int d\Omega_{\vec{k}}&A^*(\vec{p'};\vec{k})A(\vec{p};\vec{k})=
(4\pi)^2\sum_{l_1,l'_1,m_1,m'_1}\sum_{l_2,l'_2,m_2,m'_2}a^*_{l_1,m_1,l'_1,m'_1}a_{l_2,m_2,l'_2,m'_2}\\
&\times&Y^*_{l_1m_1}(\theta_{34},\phi_{34})Y_{l_2m_2}(\theta_{12},\phi_{12})\int
d\Omega_{\vec{k}}Y^*_{l'_1m'_1}(\theta_{\vec{k}},\phi_{\vec{k}})
Y_{l'_2m'_2}(\theta_{\vec{k}},\phi_{\vec{k}})\cr
&=&(4\pi)^2\sum_{l_1,l_2,l'_1,m_1,m_2,m'_1}a^*_{l_1,m_1,l'_1,m'_1}a_{l_2,m_2,l'_1,m'_1}(-1)^{m_1}Y_{l_1-m_1}(\theta_{34},\phi_{34})
Y_{l_2m_2}(\theta_{12},\phi_{12})\ ,\nonumber
\en
where we have used $Y^*_{lm}(\theta,\phi)=(-1)^mY_{l,-m}(\theta,\phi)$.
Inserting (\ref{rhs}) into (\ref{unitarity}), one obtains:
\bn
(-i)\sum_{l,l',m,m'}[a_{lm,l'm'}(s)-(-1)^{m+m'}a^*_{l',-m',l,-m}(s)]
Y_{lm}(\theta_{12},\phi_{12})Y_{l'm'}(\theta_{34},\phi_{34})\cr
=\frac{p}{4\pi\sqrt s}
\sum_{l,l',l'_1,m,m',m'_1}a^*_{l',-m',l'_1,m'_1}(s)a_{l,m,l'_1,m'_1}(s)(-1)^{m'}
Y_{lm}(\theta_{12},\phi_{12})Y_{l'm'}(\theta_{34},\phi_{34})\ .
\en
As the spherical harmonics form a complete and orthonormal set, 
the equality of the coefficients of the expansions follows and the elastic unitarity condition finally takes
the form:
\be\label{coeff}
(-i)[a_{lm,l'm'}(s)-(-1)^{m+m'}a^*_{l',-m',l,-m}(s)]=\frac{p}{4\pi\sqrt s}\sum_{l_1,m_1}(-1)^{m'}
a^*_{l',-m',l_1,m_1}(s)a_{l,m,l_1,m_1}(s)\ .
\ee

From this expression we can get the bounds on the partial wave amplitudes. 
Taking e.g. in (\ref{coeff}) $m'=m=0$ and $l=l'$, one obtains:
\bn
(-i)[a_{l0,l0}(s)-a^*_{l0,l0}(s)] &=&\frac{p}{4\pi\sqrt s}\sum_{l_1,m_1}a^*_{l0,l_1m_1}(s)a_{l0,l_1m_1}(s)\cr
&=&\frac{p}{4\pi\sqrt s}\sum_{l_1,m_1}|a_{l0,l_1m_1}(s)|^2
\geq\frac{p}{4\pi\sqrt s}|a_{l0,l0}(s)|^2\ .
\en
Thus
\be
Im\ a_{l0,l0}(s)\geq\frac{p}{8\pi\sqrt s}|a_{l0,l0}(s)|^2\ ,
\ee
which is equivalent to
\be\label{bound}
|a_{l0,l0}(s)-i\frac{8\pi\sqrt s}{p}|\leq \frac{8\pi\sqrt s}{p}\ .
\ee
This is an expression of the elastic unitarity condition for the partial-wave amplitudes
and it should be compared to the formula for the commutative case (the normalizations chosen in 
(\ref{pw_cc}) and
(\ref{pw}) correspond to each other, as $Y_{l0}(\theta,0)=\sqrt{\frac{2l+1}{4\pi}}P_l(\cos\theta)$)
\be
Im\ a_l(s)=\frac{p}{8\pi\sqrt s}|a_l(s)|^2\ .
\ee

\section{Conclusions}

We have investigated the validity of the tree-unitarity criterion \cite{{tikt1},{ls}} for quantum field
theories with space-time \ncy, treated according to the \nc\ extension of the TOPT developed in \cite{Liao1}.
We have found that the tree-unitarity condition is fullfilled by the NC $\phi^3$ scalar theory, which might have
beneficial implications for its exact unitarity and renormalizability.

We have also derived the unitarity constraint on the partial wave expansion of a $2\rightarrow 2$ scattering
amplitude in the general case of \nc\ space-time with a constant \ncy\ parameter $\theta_{\mu\nu}$, which is
an essential step in deriving Froissart-Martin-type of bounds on the cross-sections and scattering amplitudes
on \nc\ space-time.

\vskip 0.3cm
{\Large\bf{Acknowledgements}} 

Our special thanks go to Y. Liao for many enlightening discussions. We are grateful to  
M. Mnatsakanova and Yu. Vernov  for useful discussions. 

The financial support of the Academy of Finland under the Project No. 54023
is acknowledged.

\end{document}